\documentclass[12pt,a4paper]{article}
\usepackage[dvips]{graphics} 
\usepackage{cite}  
\usepackage{epsfig}
\usepackage{calc,ifthen}
\usepackage{a4p}
\newcommand{\kshort}{\ensuremath{{\rm K}^{0}_{\rm S}}}
\renewcommand{\l}{\ensuremath{\Lambda}}
\newcommand{\lb}{\ensuremath{\bar{\Lambda}}}

\newcommand{\llb}{\ensuremath{\Lambda\bar{\Lambda}}}
\renewcommand{\ll}{\ensuremath{\Lambda\Lambda}}
\newcommand{\lblb}{\ensuremath{\bar{\Lambda}\bar{\Lambda}}}
\newcommand{\llbcorr}{\ensuremath{\Lambda\bar{\Lambda}_{\rm corr}}}
\newcommand{\llbb}{\ensuremath{\Lambda\Lambda(\bar{\Lambda}\bar{\Lambda})}}
                          
\newcommand{\epem}{\ensuremath{{\rm e}^+{\rm e}^-}}

\newcommand{\zzero}{\ensuremath{{\rm Z}^0}}

\newcommand{\dy}{\ensuremath{|\Delta y|}}
\newcommand{\thet}{\ensuremath{\theta^*}}
\newcommand{\costhet}{\ensuremath{\cos \theta^*}}

\newcommand{\durh}{\ensuremath{{\rm k}_{\perp}}}
\newcommand{\jt}  {{\sc Jetset}}
\newcommand{\hw}  {{\sc Herwig}}
\newcommand{\mops}{{\sc Mops}}
\newcommand{\lep} {{\sc Lep}}
\newcommand{\opal}{{\sc Opal}}
\renewcommand{\aleph}{{\sc Aleph}}
\newcommand{\delphi}{{\sc Delphi}}

\newcommand{\beq}{\begin{equation}}
\newcommand{\eeq}{\end{equation}}

\def\lamplace{\rule{0cm}{2.5ex}}

\newcommand{\refnum} {CERN-EP/98-114}
\newcommand{\Date} {July 16, 1998}
\newcounter{hours}\newcounter{minutes}

\newcommand{\Printtime}{%
  \setcounter{hours}{\time/60}%
  \setcounter{minutes}{\time-\value{hours}*60}%
  \ifthenelse{\value{hours}<10}{0}{}\thehours:%
  \ifthenelse{\value{minutes}<10}{0}{}\theminutes}
\begin{document}
%
%  Title Page
\begin{titlepage}
\setcounter{page}{0}
%     Header
%
\vspace*{-2cm}
\begin{center}{\large EUROPEAN LABORATORY FOR PARTICLE PHYSICS}
\end{center} \bigskip
\begin{flushright}
%    \large
    \refnum  \\  \Date
\end{flushright}
%\vskip 1.cm
\bigskip\bigskip\bigskip\bigskip\bigskip
%\begin{center}
%{\Huge  --  F I N A L   DRAFT --}
%\end{center}

 \begin{center} \huge\bf\boldmath
  A Study of Parton Fragmentation \\
  in Hadronic \zzero\ Decays\\
  Using \llb\ Correlations
\end{center}\bigskip \bigskip

\begin{center}{\LARGE The OPAL Collaboration}
\end{center} \bigskip\bigskip\bigskip

\begin{center}{\large  Abstract}\end{center}
%\begin{abstract}
The correlated production of \l\ and \lb\ baryons has been studied using 
4.3 million multihadronic \zzero\ decays recorded with the \opal\ detector at 
\lep . 
Di-lambda pairs were investigated in the full data sample and for the first 
time also in 2-jet and 3-jet events selected with the \durh\ algorithm. 
The distributions of rapidity differences from correlated \llb\ pairs
exhibit short-range, local correlations and prove to be a sensitive
tool to test models, particularly for 2-jet events.
The \jt\ model describes the data best but some extra parameter tuning is 
needed to improve agreement with the experimental results  in 
the rates and the rapidity spectra simultaneously.
The recently developed modification of \jt, the MOdified Popcorn Scenarium 
(\mops), and also \hw\ do not give satisfactory results. 
This study of di-lambda production in 2- and 3-jet events supports the 
short-range compensation of quantum numbers.
%\end{abstract}

\bigskip \bigskip \bigskip \bigskip \bigskip
 
%\begin{center}
%{\bf Authors}: B. Nellen, R. Schmitz, E. von Toerne
%\end{center}
%\begin{center}
%{\bf Editorial board}:  Richard Hemingway, George Lafferty, Alfred Lee, 
%                        James Letts
%\end{center}

%\begin{center}
%{\bf Comments} to Ruth.Schmitz@cern.ch by {\bf Wednesday, July 15, 1998, 
%noon}, please. 
%\end{center}  

\begin{center}{\large (Submitted to Physics Letters B)}
\end{center}

% \begin{center}
% {\large Submitted to Zeitschrift f\"ur Physik C }
% \end{center}
%This version produced at \Printtime, on \today
\end{titlepage}
%
%Author-List:
\begin{center}{\Large        The OPAL Collaboration
}\end{center}\bigskip
\begin{center}{
%begin authorlist PLEASE DO NOT DELETE THIS COMMENT
G.\thinspace Abbiendi$^{  2}$,
K.\thinspace Ackerstaff$^{  8}$,
G.\thinspace Alexander$^{ 23}$,
J.\thinspace Allison$^{ 16}$,
N.\thinspace Altekamp$^{  5}$,
K.J.\thinspace Anderson$^{  9}$,
S.\thinspace Anderson$^{ 12}$,
S.\thinspace Arcelli$^{ 17}$,
S.\thinspace Asai$^{ 24}$,
S.F.\thinspace Ashby$^{  1}$
% ...
%end authorlist PLEASE DO NOT DELETE THIS COMMENT
}\end{center}\bigskip
\bigskip
%begin institutes
$^{  1}$School of Physics and Astronomy, University of Birmingham,
Birmingham B15 2TT, UK
\newline
$^{  2}$Dipartimento di Fisica dell' Universit\`a di Bologna and INFN,
I-40126 Bologna, Italy
\newline
$^{  3}$Physikalisches Institut, Universit\"at Bonn,
D-53115 Bonn, Germany
\newline
$^{  4}$Department of Physics, University of California,
Riverside CA 92521, USA
\newline
$^{  5}$Cavendish Laboratory, Cambridge CB3 0HE, UK
\newline
$^{  6}$Ottawa-Carleton Institute for Physics,
Department of Physics, Carleton University,
Ottawa, Ontario K1S 5B6, Canada
\newline
$^{  7}$Centre for Research in Particle Physics,
Carleton University, Ottawa, Ontario K1S 5B6, Canada
\newline
$^{  8}$CERN, European Organisation for Particle Physics,
CH-1211 Geneva 23, Switzerland
\newline
$^{  9}$Enrico Fermi Institute and Department of Physics,
University of Chicago, Chicago IL 60637, USA
\newline
$^{ 10}$Fakult\"at f\"ur Physik, Albert Ludwigs Universit\"at,
D-79104 Freiburg, Germany
\newline
$^{ 11}$Physikalisches Institut, Universit\"at
Heidelberg, D-69120 Heidelberg, Germany
\newline
$^{ 12}$Indiana University, Department of Physics,
Swain Hall West 117, Bloomington IN 47405, USA
\newline
$^{ 13}$Queen Mary and Westfield College, University of London,
London E1 4NS, UK
\newline
$^{ 14}$Technische Hochschule Aachen, III Physikalisches Institut,
Sommerfeldstrasse 26-28, D-52056 Aachen, Germany
\newline
$^{ 15}$University College London, London WC1E 6BT, UK
\newline
$^{ 16}$Department of Physics, Schuster Laboratory, The University,
Manchester M13 9PL, UK
\newline
$^{ 17}$Department of Physics, University of Maryland,
College Park, MD 20742, USA
\newline
$^{ 18}$Laboratoire de Physique Nucl\'eaire, Universit\'e de Montr\'eal,
Montr\'eal, Quebec H3C 3J7, Canada
\newline
$^{ 19}$University of Oregon, Department of Physics, Eugene
OR 97403, USA
\newline
$^{ 20}$CLRC Rutherford Appleton Laboratory, Chilton,
Didcot, Oxfordshire OX11 0QX, UK
\newline
$^{ 22}$Department of Physics, Technion-Israel Institute of
Technology, Haifa 32000, Israel
\newline
$^{ 23}$Department of Physics and Astronomy, Tel Aviv University,
Tel Aviv 69978, Israel
\newline
$^{ 24}$International Centre for Elementary Particle Physics and
Department of Physics, University of Tokyo, Tokyo 113, and
Kobe University, Kobe 657, Japan
\newline
$^{ 25}$Institute of Physical and Environmental Sciences,
Brunel University, Uxbridge, Middlesex UB8 3PH, UK
\newline
$^{ 26}$Particle Physics Department, Weizmann Institute of Science,
Rehovot 76100, Israel
\newline
$^{ 27}$Universit\"at Hamburg/DESY, II Institut f\"ur Experimental
Physik, Notkestrasse 85, D-22607 Hamburg, Germany
\newline
$^{ 28}$University of Victoria, Department of Physics, P O Box 3055,
Victoria BC V8W 3P6, Canada
\newline
$^{ 29}$University of British Columbia, Department of Physics,
Vancouver BC V6T 1Z1, Canada
\newline
$^{ 30}$University of Alberta,  Department of Physics,
Edmonton AB T6G 2J1, Canada
\newline
$^{ 31}$Duke University, Dept of Physics,
Durham, NC 27708-0305, USA
\newline
$^{ 32}$Research Institute for Particle and Nuclear Physics,
H-1525 Budapest, P O  Box 49, Hungary
\newline
$^{ 33}$Institute of Nuclear Research,
H-4001 Debrecen, P O  Box 51, Hungary
\newline
$^{ 34}$Ludwigs-Maximilians-Universit\"at M\"unchen,
Sektion Physik, Am Coulombwall 1, D-85748 Garching, Germany
\newline
%end institutes
\bigskip\newline
%begin notes
$^{  a}$ and at TRIUMF, Vancouver, Canada V6T 2A3
\newline
$^{  b}$ and Royal Society University Research Fellow
\newline
$^{  c}$ and Institute of Nuclear Research, Debrecen, Hungary
\newline
$^{  d}$ and Department of Experimental Physics, Lajos Kossuth
University, Debrecen, Hungary
\newline
$^{  e}$ on leave of absence from the University of Freiburg
\newline
%end notes

\newpage

%\input{introduction}
%
%... CORRPAPER ... introduction.tex   /  update: 09.07.98 RES
%
\section{Introduction}
\label{sec:intro}
The compensation of quantum numbers plays a key role in our understanding of the
fragmentation process whereby partons transform into observable hadrons. 
Consequently, baryon production in hadronic \epem\ annihilation final states
provides data very well suited to test 
phenomenological fragmentation models. In particular, the study of di-lambda 
pairs allows a subtle testing of model predictions because of the relatively 
large rates and the necessity to compensate two quantum numbers:  
baryon number and strangeness.  

Fragmentation models such as \jt \cite{jt} and \hw \cite{hw} are based on
a chainlike production of hadrons with local compensation of quantum numbers.
In \jt, particle production is implemented via string fragmentation. Baryons (B)
are formed when a diquark pair is contained in the string 
%(figure~\ref{fig:mops}a)
(see diagram a below), thus resulting in a strong baryon-antibaryon 
correlation. This correlation can be softened by the ``popcorn effect'' when 
an additional meson (M) is produced between the baryon pair as shown in 
%figure~\ref{fig:mops}b,c. 
the diagrams b and c below.
In contrast, \hw\ describes fragmentation via the formation of clusters and 
their subsequent decay. Baryons are produced by the isotropic cluster decay 
into a baryon pair, which can result in stronger correlations than those 
predicted by \jt .

\vspace*{-2.5cm}
\begin{center}
  \resizebox{\textwidth}{!}{
    \includegraphics{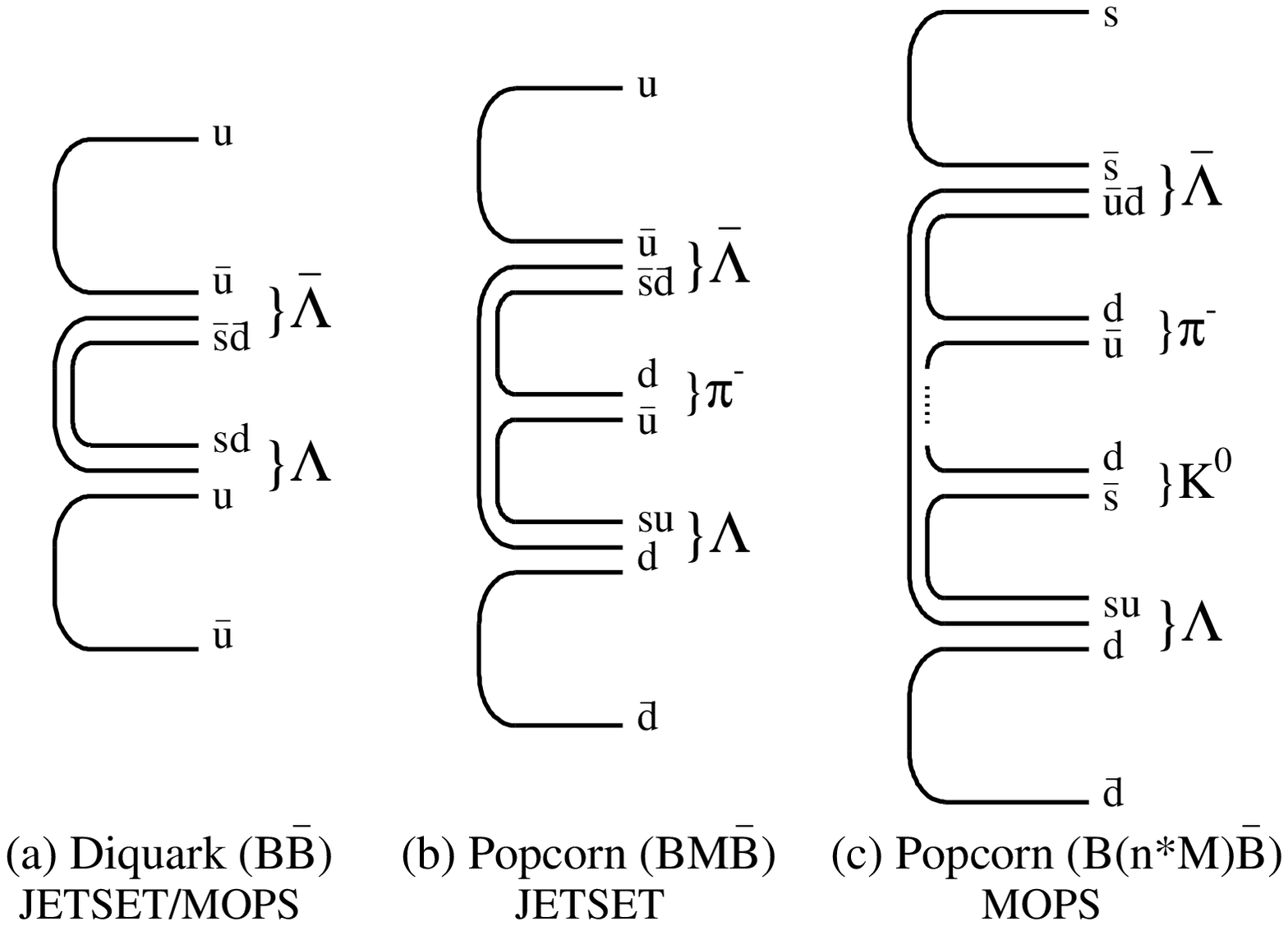}
    }
\end{center}
\vspace{-2.5cm}

Di-lambda production in multihadronic \zzero\ decays has been studied over the 
past years by experiments at {\sc Petra}, {\sc Pep} and \lep 
\cite{petra,pep,aleph_corr,delphi_corr,opal_corr}. These experiments 
report short-range correlations as observed in the distributions of the 
rapidities $y$ or rapidity differences \dy\ of correlated \llb\ pairs.
The rapidity of a particle is defined as
$y=\frac{1}{2}\ln\left(\frac{E+p_{\scriptscriptstyle\parallel}}{E-
p_{\scriptscriptstyle\parallel}}\right)$, 
where $E$ is the energy of the particle and $p_{\scriptscriptstyle\parallel}$ 
the longitudinal momentum with respect to the thrust axis. Rapidity differences 
are Lorentz-invariant under boosts along the event axis.
These correlations are compared to predictions of \jt\ and \hw . 
Satisfactory agreement is found with the predictions of \jt\ when $\rho$, the 
``popcorn parameter''\footnote{The value of $\rho$ can be set in \jt\ with the 
parameter PARJ(5): $\rho=\frac{BM\bar{B}\lamplace}{B\bar{B}+BM\bar{B}} = 
\frac{\rm PARJ(5)}{0.5+{\rm PARJ(5)}}$.},
is set to the default value, $\rho=0.5$. However, the tune of other parameters 
modeling baryon production significantly influences the predictions 
\cite{james_early}. 
\hw\ on the other hand predicts correlations much larger than those 
experimentally observed.

The full data sample of 4.3~million hadronic \zzero\ decays collected with 
the \opal\ detector at \lep\ in the region of the \zzero\ peak is used in this 
investigation. It supplements the earlier \opal\ 
work\cite{opal_corr} by increased statistics and a more robust technique  
to remove the background contributions from the \llb\ and \llbb\ 
samples in order to obtain a correlated \llb\ sample which is as clean as 
possible. 
The \llb\ correlations are investigated mainly via rapidity differences. 
They are compared to the earlier \lep\ results and to the predictions of  
\jt\ and \hw. The predictions of the recent \jt\ modification 
\mops\ (MOdified Popcorn Scenarium)~\cite{mops} are also considered. 
Correlated \llb\ pairs are also studied in 2-jet events in which models can
be tested with improved sensitivity (compared to the full data sample)
when rapidity differences are investigated. 
Finally, we study correlated \llb\ pairs within the same and within different
jets. 

Section~\ref{sec:procedure} gives a short description of the \opal\ detector 
and
presents the selection of the \l\ events\footnote 
{For simplicity \l\ refers to both \l\ and \lb.}
in the total sample and also in 2-jet and 3-jet events, for both experimental 
and simulated data.
In section~\ref{sec:method} the separation of the  \llb\ and  \llbb\ samples 
from the background and the determination of the rates of correlated 
\llb\ pairs as a function of the rapidity differences \dy\ are discussed. 
Section~\ref{sec:results} contains the measured rates with their errors and a 
comparison to earlier results as well as the presentation of the differential 
distributions as a function of \dy\ and \costhet, where \thet\  is the angle 
between the thrust axis and the \l\ momentum calculated in the rest frame of 
the di-lambda pair.
In section~\ref{sec:comparison} the models are tested using the production 
rates of \l\ pairs as well as the \costhet\ and \dy\ spectra of correlated 
\llb\ pairs. 
The range of di-lambda correlations is investigated in section~\ref{sec:jets}
by the assignment of the \l's to the jets. 
Conclusions are drawn in section~\ref{sec:summary}.
%
%
%\input{procedure}
%
%... corrpaper ... PROCEDURE.TEX   /  update: 8-JUL-98 RES
%
\section{Experimental Procedure}
\label{sec:procedure}
\subsection{The OPAL Detector}
\label{subs:detec}
A detailed description of the \opal\ detector can be found in 
Ref.~\cite{detector}. 
Of most relevance for the present analysis is the tracking system and 
the electromagnetic calorimeter. 
The tracking system consists of a silicon microvertex detector, an inner
vertex gas chamber, a large-volume jet chamber and specialized chambers at
the outer radius of the jet chamber which improve the measurements in
the $z$ direction ($z$-chambers)\footnote
{The coordinate system is defined so that $z$~is the coordinate parallel 
to the e$^-$ beam axis, $r$~is the coordinate normal to the beam axis, 
$\phi$~is the azimuthal angle around the beam axis, and $\theta$~is the 
polar angle \mbox{with respect to~$z$.}}. The tracking system covers the region
$|\cos\theta|<0.95$ and is located within a solenoidal magnet coil with
an axial field of~0.435~T. 
The tracking detectors provide momentum measurements of charged
particles, and particle identification from measurements of the
ionization energy loss,  d$E$/d$x$.
Electromagnetic energy is measured by a lead-glass calorimeter 
located outside the magnet coil, which covers $|\cos\theta|<0.98$.
\subsection{Data Samples}
\label{subs:data}
The analysis is based on hadronic \zzero\ decays collected around the 
\zzero\ peak from 1990 to 1995 (total \lep\ 1 statistics). The hadronic events 
were selected with the standard \opal\ procedure~\cite{tkmh} based on the 
number and quality of the measured tracks and the electromagnetic clusters and 
on the amount of visible energy in the event. In addition, events with the 
thrust axis close to the beam direction were rejected by requiring 
$|\cos \theta_{\rm {thrust}}|< 0.9$, where $\theta_{\rm thrust}$ is the polar 
angle of the thrust axis. With the additional requirement that the jet chamber
and the z-chambers were fully operational, a total of 3.895 million hadronic 
events remained for further analysis, with an efficiency of ($98.4 \pm 0.4$)\%. 
The remaining background processes, 
such as $\epem \rightarrow \tau^+ \tau^-$ and two photon events, were estimated 
to be at negligible level (0.1\% or less).

After the \l-selection which will be described below, the selection of 2- and 
3-jet events was performed. Charged tracks and electromagnetic 
clusters not associated with any track were grouped into jets using the 
\durh\ recombination algorithm \cite{durham} with a cut value
$y_{\rm cut}=0.005$. In addition to the standard selection criteria, the
energy of the clusters and the momenta of the charged tracks had to be less 
than 60~GeV/$c$.
To improve the quality of the jets it was finally required that there be at 
least two charged particles per jet (in addition to the possible tracks from 
\l\ decays) and that the minimum energy per jet was $5$~GeV.
The cuts on the quality of jets were chosen to be this loose to keep the 
kinematic range as large as possible for comparison with fragmentation models. 
In total, samples of 1.7 million 2-jet events and 1.4 million 3-jet events 
were available for further analysis corresponding to $45\%$ and $36\%$,
respectively, of the entire data set.
\subsection{Monte Carlo Event Samples} 
\label{subs:mc} 
Monte Carlo hadronic events with a full simulation of the OPAL detector 
\cite{gopal} and including initial-state photon radiation were used 
(a) for evaluation of detector acceptance and resolution 
and (b) for studying the efficiency of the di-lambda reconstruction as a 
function of the rapidity differences.  
In total, seven million simulated events were available, of which four million
were generated by \jt~7.4 with fragmentation parameters  
described in~\cite{jt7.4}, and three million were generated by \jt~7.3  with  
fragmentation parameters described in~\cite{jt7.3}.
The two \jt\ versions differ in the particle decay tables and heavy meson  
resonances. 
There are also some differences in the simulation of baryon production between 
the two samples.
Their small influence on the efficiency correction to the experimental 
data is accounted for in the systematic errors (see 
section~\ref{subs:syserr}). 

For comparison with the experimental results, the Monte Carlo models 
\jt~7.4 and \hw~5.9\cite{hw5.8}\footnote 
{The fragmentation parameters of \hw~5.9 were identical to those used in  
 our tuned version of \hw~5.8\cite{hw5.8} with the exception of the maximum
 cluster mass ({\tt CLMAX}) which was set to 3.75 GeV in order to improve the  
 description of the mean charged particle multiplicity in inclusive  
 hadronic \zzero\ decays.} 
were used. Both models give a good description of global  
event shapes and many inclusive particle production rates, but differ in  
their description of the perturbative phase and their implementation of the  
hadronization mechanism. 

Tracks and clusters are selected in the Monte Carlo events, which 
include detector simulation, in the same way as for the data, and the 
resulting four-vectors of particles are referred to as being at the 
`detector level'.  Alternatively, for testing the model predictions, 
Monte Carlo samples without 
initial-state photon radiation nor detector simulation are used, with 
all charged and neutral particles with mean lifetimes greater than 
$3\times10^{-10}$~s treated as stable.  The four-vectors of the 
resulting particles are referred to as being at `generator level'. 
\subsection{\l\ Reconstruction } 
\label{subs:lambda} 
Neutral strange \l\ baryons were reconstructed in their decay  
channel \l ~$\rightarrow \pi^-$p as described in \cite{opal_sp}. 
Briefly, tracks of opposite charge were paired and regarded as a secondary  
vertex candidate if the track pair intersection in the plane  
perpendicular to the beam axis satisfied the criteria of a neutral two-body  
decay with a decay length of at least 1 cm.
 
Each candidate track pair was refitted with the 
constraint that the tracks originated from a common vertex, and 
background from photon conversions was suppressed. Information from  
d$E$/d$x$ measurements was used as in \cite{opal_sp} to help identify the 
$\pi$ and p for further background suppression, primarily due to 
\kshort $\rightarrow \pi^+ \pi^-$. Two sets of cuts, called `method 1' and 
`method 2' are described in \cite{opal_sp} for \l\ identification.   
For the present analysis, \l\ candidates were reconstructed using method~1,  
which is optimized to have good mass and momentum resolution. 

\begin{figure}[!htb]
  \begin{center}
    \resizebox{\textwidth}{!}{
      \includegraphics{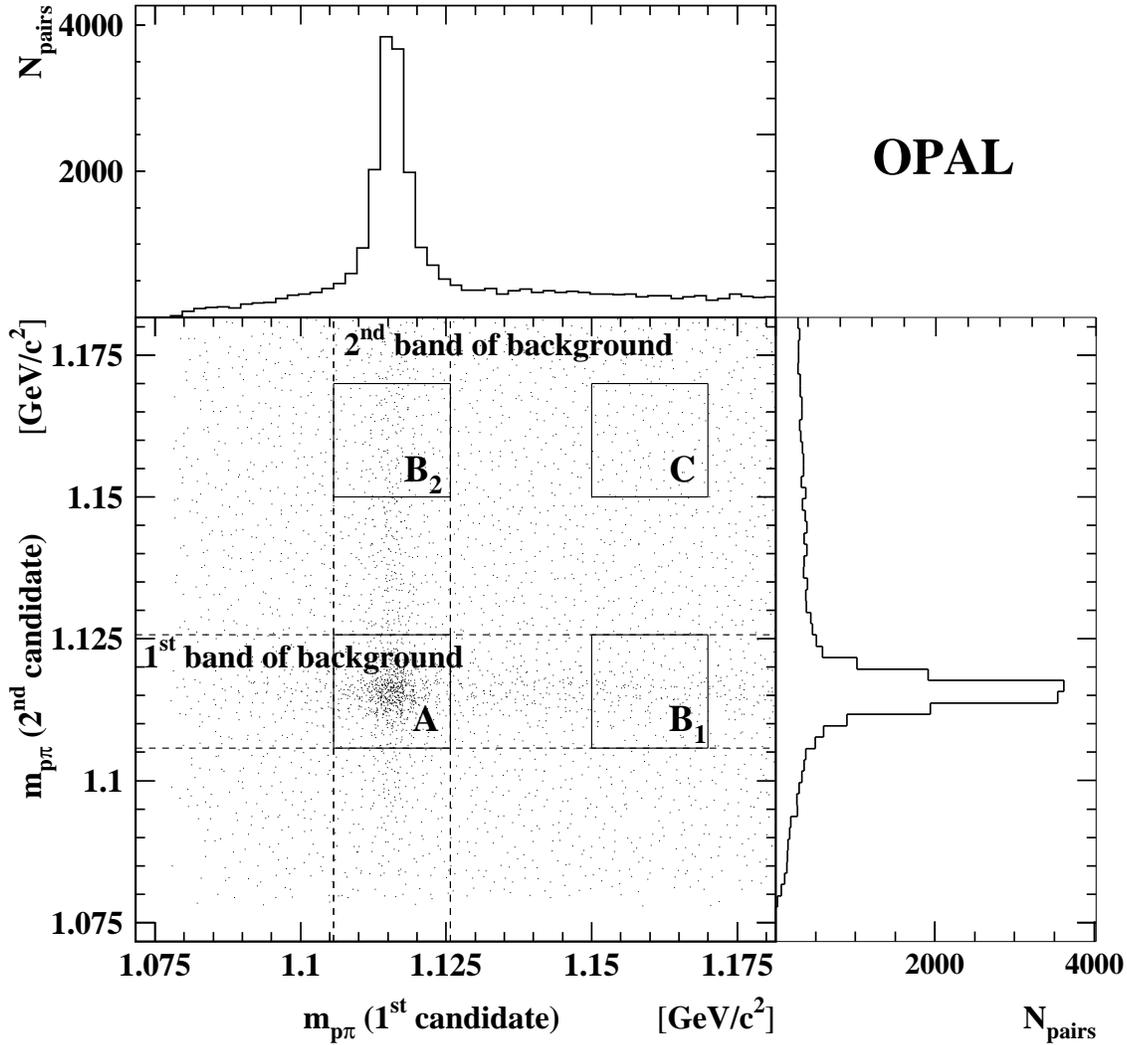}
      }
  \end{center}
  \vspace{-1.cm.}
  \caption{Two-dimensional mass distribution of \llb\ candidates and 
    projections onto the mass axes. The background forms a horizontal and a 
    vertical band from pairs with one fake \l\ above a uniform background from
    two non-\l\ candidates. The signal peak at the \l\ mass of 1.116~GeV/$c^2$
    is clearly visible.}
  \label{fig:m2dim}
\end{figure}

\begin{sloppypar} 
By these means a narrow \l\ mass peak above a small background has been 
obtained. The selection of di-lambda candidates with both invariant masses in  
the range \mbox{1.1057~GeV/$c^2<$ $m_{\pi p}<$ 1.1257~GeV/$c^2$} (region A in 
figure~\ref{fig:m2dim}) retains most of the \l\ signal for further analysis. 
\end{sloppypar}
%
%
%\input{method}
%
%... corrpaper ... METHOD.TEX   /  update: 09-JUL-98 RES
%
\section{Selection of Correlated \l-pairs}
\label{sec:method}
\subsection{Method}
\label{subs:mgeneral}
Events with more than one \l\ candidate that had passed the above selection 
criteria were considered and all possible pair combinations of the \l\ and 
\lb\ baryons within an event were formed. This resulted in pairs of \llb, 
\l\l\ and \lb\lb. Combinations were rejected if the pair had a track in 
common. The remaining pairs are henceforth referred to as \l-pair candidates. 

The three types of baryon pairs can be grouped into two classes: pairs with 
different baryon numbers \llb\ and pairs with equal baryon numbers \llbb. 
Only in \llb\ pairs can the baryon and flavor quantum numbers be compensated 
by correlated production. \llbb\ pairs can never be produced in correlation
and hence they will occur only in events with more than one baryon-antibaryon 
pair (B$\bar{\rm B}$). In such events uncorrelated \llb\ pairs from 
different (B$\bar{\rm B}$) pairs are also possible. The number of uncorrelated 
\llb\ pairs corresponds to the number of pairs with same baryon number. 
Hence, the number of correlated \llb\ pairs can be derived via 
\beq
N_{\llb}^{\rm corr.} = N_{\llb} - (N_{\ll} + N_{\lblb})\: .
\eeq
At this stage 9479 \llb\ and 4217 (\l\l+\lb\lb) pair candidates are 
selected.
\subsection{Background Subtraction and Efficiency Correction}
\label{subs:backgr} 
Due to the small statistical errors it is necessary to keep systematic 
uncertainties as low as possible in this analysis. 
The correct subtraction of {\mbox{non-\l\ }} background from the pairs is 
therefore of  particular importance. This background consists mainly of other 
long-lived particles with similar decay topologies (namely \kshort $\rightarrow
\pi^+ \pi^-$) and random 
track  combinations. An important contribution to the contamination is the 
so-called correlated background from \l candidates that have been reconstructed 
with one false decay track.   
They are more numerous in pairs with opposite baryon number because the 
number of \llb\ pairs is far higher than the number of \llbb\ pairs.  
For this reason the background has to be estimated in the two samples  
separately. 
Background pairs occur when either one or both \l-candidates are fake.
In the two-dimensional mass plane in figure~\ref{fig:m2dim}, pairs with one 
fake \l\ form horizontal and vertical bands of background, while pairs with 
two fake candidates are uniformly distributed in the region above the lower 
mass bounds. 

The background was subtracted using a two-dimensional sideband method. The  
background in the signal region A was measured from two mass windows 
(sidebands) of the same size (regions B$_1$ and B$_2$) placed 
in the two bands of background.  
In this way the background with two fake candidates is counted twice. The latter
was determined from region C outside the bands.
Hence, the signal is obtained with the subtraction:  
\begin{center} 
   Signal = $N_{\rm A} - (N_{\rm B_1} + N_{\rm B_2} - N_{\rm C})$ .   
\end{center}
We optimized the position of the sidebands with a MC test investigating the
deviations between the background-corrected sample and the true-\l\ sample.  
The stability of this method was tested in the experimental data by shifting 
the position of the sidebands by one half of the band size from the optimized 
position. The fluctuations were of the same size as the deviations found in 
the MC. 

Finally the background-corrected \llb\ and \llbb\ signal distributions were 
corrected for detector acceptance and reconstruction efficiency as functions 
of \dy\ and \costhet . The average efficiency in the total hadronic sample 
was found to be $\approx 2\%$, varying between 1.3\% and 2.5\% over the 
\dy/\costhet\ range.
%
%
%\input{results}
%
%... CORRPAPER ... results.tex   /  update: 08-JUL-98 RES
%
\section{Experimental Results}
\label{sec:results}
In an earlier OPAL paper~\cite{opal_corr} based on the 1990 and 1991 data 
samples we already investigated the production dynamics of baryon-antibaryon 
pairs. 
In this section, we present the rates and differential distributions of 
\l~pairs using the full 1990 to 1995 LEP~1 data in three samples: 
the entire set of multihadronic events, the 2-jet and the 3-jet events. 
\subsection{Pair Production Rates}
\label{subs:rates}
The resulting rates for \llb\ and \llbb\ pairs in all hadronic events, 
determined as sum over all corrected \dy\ bins, are 
given in table~\ref{tab:rates}. The rates for the correlated \llb\ pairs 
are derived according to equation (1) from the difference of the opposite 
and same baryon number pairs. Compared to the results from other \lep\ 
experiments and to the previous \opal\ publication, good agreement is found. 

The di-lambda rates in 2- and 3-jet events are listed in  
table~\ref{tab:2_3jrates}.
In 3-jet events, due to the higher color charge of the gluons, the average 
pair multiplicity is higher. 
\subsection{Differential Distributions}
\label{subs:distributions}
\begin{figure}[!htb]
  \vspace{-0.7cm}
  \begin{center}
    \resizebox{\textwidth}{!}{
      \includegraphics{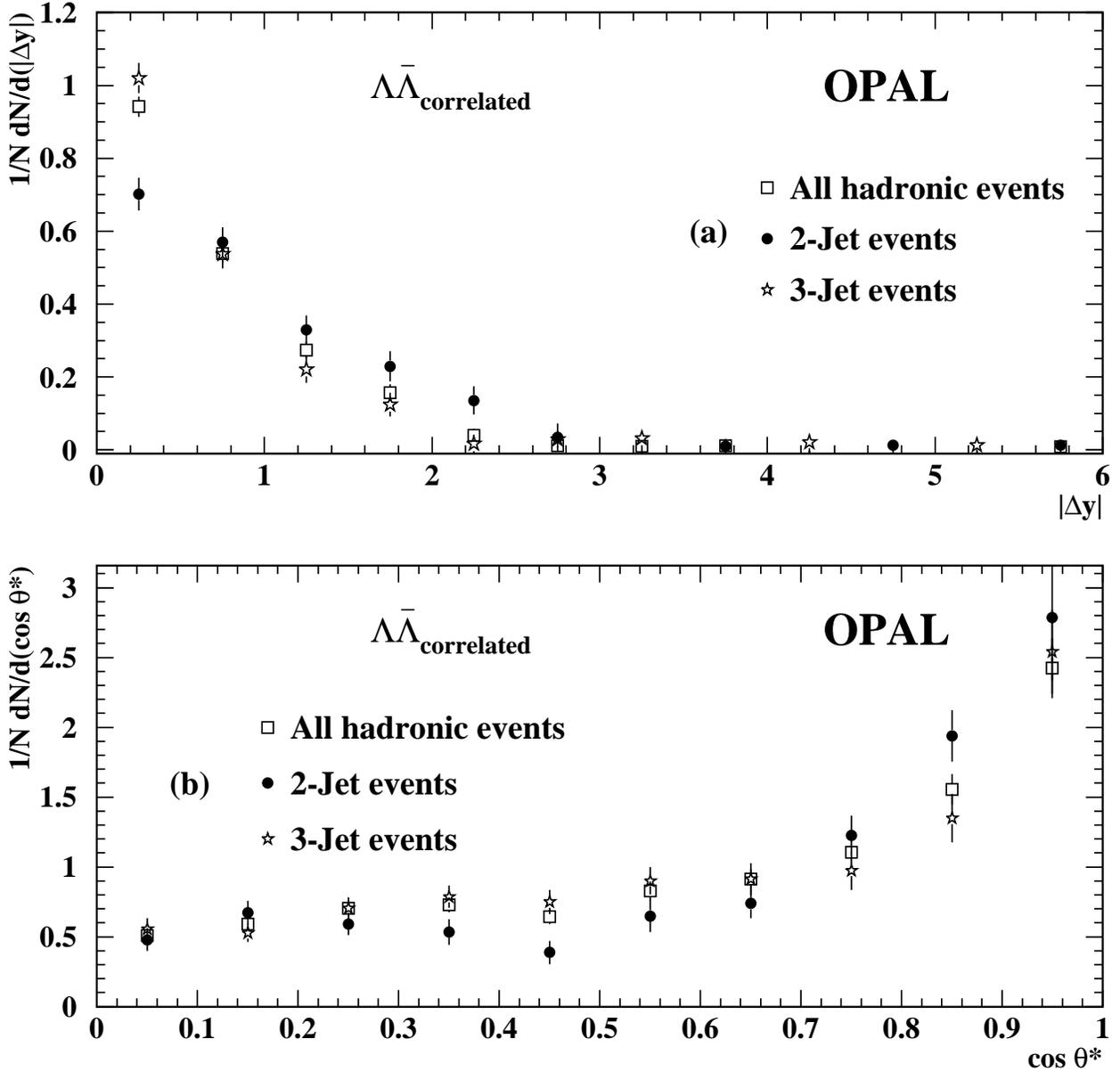}
      }
  \end{center}
  \vspace{-0.5cm}
  \caption{Comparison of the shape of differential distributions in all 
    hadronic events and in 2-/ 3-jet events for the \dy\ distribution in 
    (a) and the \costhet\ distribution in (b). The errors shown are purely 
    statistical, the influence of the systematic errors is negligible.}
   \label{fig:distrib}
\end{figure}

We studied the correlations in the differential \llb\ spectra using the 
observables \dy\ and \costhet\ as they are particularly sensitive for  
comparison with Monte Carlo models.
The differential distributions are shown in figure~\ref{fig:distrib}.
The short-range correlations show up as a peak in the region \dy\ $\le$ 2.0.

When investigating \dy\ distributions, we will restrict ourselves to 
2-jet events.
This is due to the fact that in 3-jet events many particle momenta have large 
angles to the thrust axis, resulting in smaller longitudinal momenta and 
smaller rapidity differences, independent of correlations. As a result the \dy\ 
distribution is broader and less steep in 2-jet events than in the 3-jet or the 
total sample (see figure~\ref{fig:distrib}a). Consequently, also the range of 
variations is larger in 2-jet events and yields a higher sensitivity in the 
comparison with model predictions.
\subsection{Systematic Errors}
\label{subs:syserr}
The systematic error is found to be largely independent of \dy\ and \costhet ,  
and in the subsequent discussion of the differential distributions of 
the correlated pairs, only normalized distributions are considered. These are 
largely insensitive to effects of systematic uncertainties. Consequently, the 
systmatic errors discussed below are mainly relevant for the total rates.

For the determination of the experimental uncertainties we considered the 
following sources of systematic effects:
\begin{itemize}
\item Uncertainties due to the subtraction of background via the 
  sidebands. These were estimated using simulated events by applying the 
  analysis to the fully detector simulated MC and comparing the rate from the 
  background corrected sample to the true number.
\item Efficiency uncertainties. These were estimated from the difference of 
  the results when the efficiency
  correction was done using both \jt\ versions 7.3 
  and 7.4 samples in combination and using them separately.
\item The statistical error of the efficiency due to the limited sample size
  of the simulated events at detector level.
\item Uncertainties in the modelling of the cut variables used for the \l\
  selection. This error is taken from a former analysis 
  \cite{opal_sp} where it was determined very precisely for single \l 's. 
  The error given there is doubled for the \l\ pairs in the present 
  analysis.
\end{itemize}

These effects contribute to the total systematic error as shown in
table~\ref{tab:syserr}, where the relative systematic errors from the 
different sources are compared to the total systematic as well as to the 
statistical error. Statistical and total systematic errors contribute about 
equally.
%
%
%\input{comparison}
%
%... CORRPAPER ... comparison.tex   /  update: 08-JUL-98 RES
%
\section{Comparison with Fragmentation Models}
\label{sec:comparison}

We start the discussion with the numbers and distributions of the models with
OPAL default tunes that optimize the general performance of the models and the 
agreement with the measured single particle rates. 

\subsection{Pair Production Rates}
\label{subs:comp_rates}

We investigate the di-lambda rates first in the total hadronic data sample
comparing the measured rates to the predictions of the models \jt~7.4, \mops\ 
and \hw~5.9 (see table~\ref{tab:rates}). None of the models gives a perfect 
description of the data but \hw\ clearly exhibits the largest disagreement. 

The comparison of the di-lambda rates in 2- and 3-jet events is given in 
table~\ref{tab:2_3jrates}. The higher multiplicity in 3-jet events compared to 
2-jet events is qualitatively well described by all three models. 
However, only \jt\ yields a prediction compatible with the measured numbers.
In the 2-jet event sample the agreement is excellent.
In the 3-jet sample all the measured rates exceed the \jt\ predictions. 
This can be compared to the observation that \l\ rates in gluon jets are too low
in \jt~\cite{opal_jets} .

\subsection{Differential Distributions}
\label{subs:comp_distributions}

To further investigate the nature of the \llb\ correlations we 
compare the differential distributions of correlated \llb\ pairs with the 
predictions of the various models. We use the variables \costhet\ and \dy\ 
and test their sensitivity to distinguish between the different fragmentation 
models and baryon production mechanisms. All distributions are of the type 
$ \frac{1}{N} \frac{dN}{d(\dy)}$, $N$ being the total number of entries. 
This has the advantage that they are independent of the total rates and 
that the systematic errors mostly cancel out, since they are nearly independent 
of both \dy\ and \costhet .

\begin{figure}[!htb]
  \vspace{-0.7cm}
  \begin{center}
    \resizebox{\textwidth}{!}{
      \includegraphics{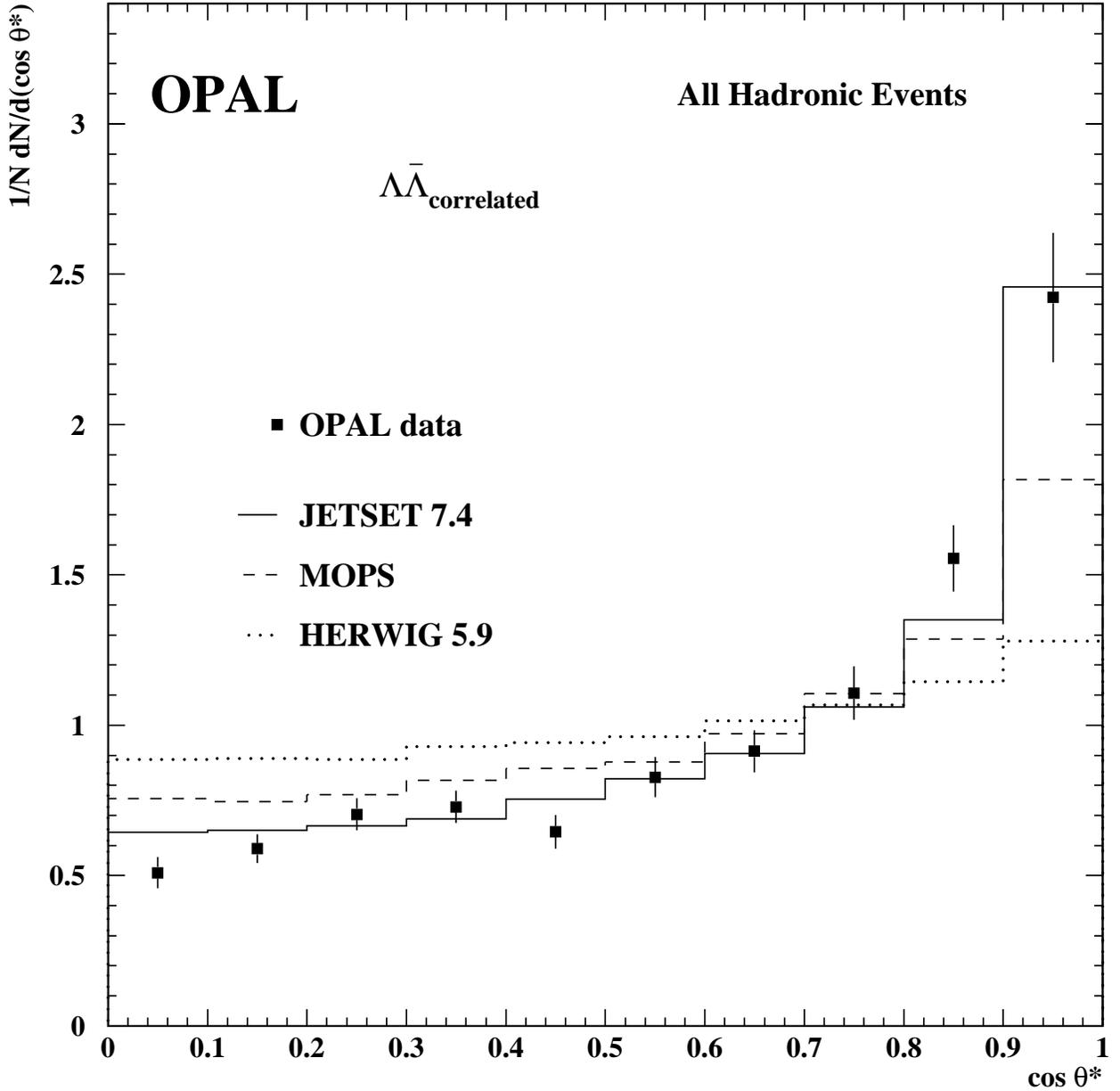}
      }
  \end{center}
  \vspace{-0.5cm}
  \caption{Comparison of the measured distribution of the angle \thet\
     for correlated \llb\ pairs in all hadronic events with the 
     predictions from the various models. The errors shown are 
     purely statistical, the influence of the systematic errors is 
     negligible.}
  \label{fig:comp_theta}
\end{figure}

The angle \thet\ is particularly suited to distinguish between string and 
cluster fragmentation.  The mostly isotropic cluster decay (\hw) results in a 
relatively flat \costhet\ distribution  whereas string fragmentation produces 
the correlated \llb\ system predominantly close to the thrust axis, i.e., with 
$\costhet \approx 1$. These predictions are compared to the measurement 
in figure~\ref{fig:comp_theta}. The data show a distribution that is strongly 
peaked towards \costhet = 1 and therefore clearly rule out the \hw\ cluster 
model. 
The predictions of \mops\ agree somewhat better with the experimental 
distribution but they also fail to model the forward peak correctly. 
Only \jt\ yields a good description of the data.

\begin{figure}[p]
  \vspace{-0.5cm}
  \begin{center}
    \resizebox{\textwidth}{!}{
      \includegraphics{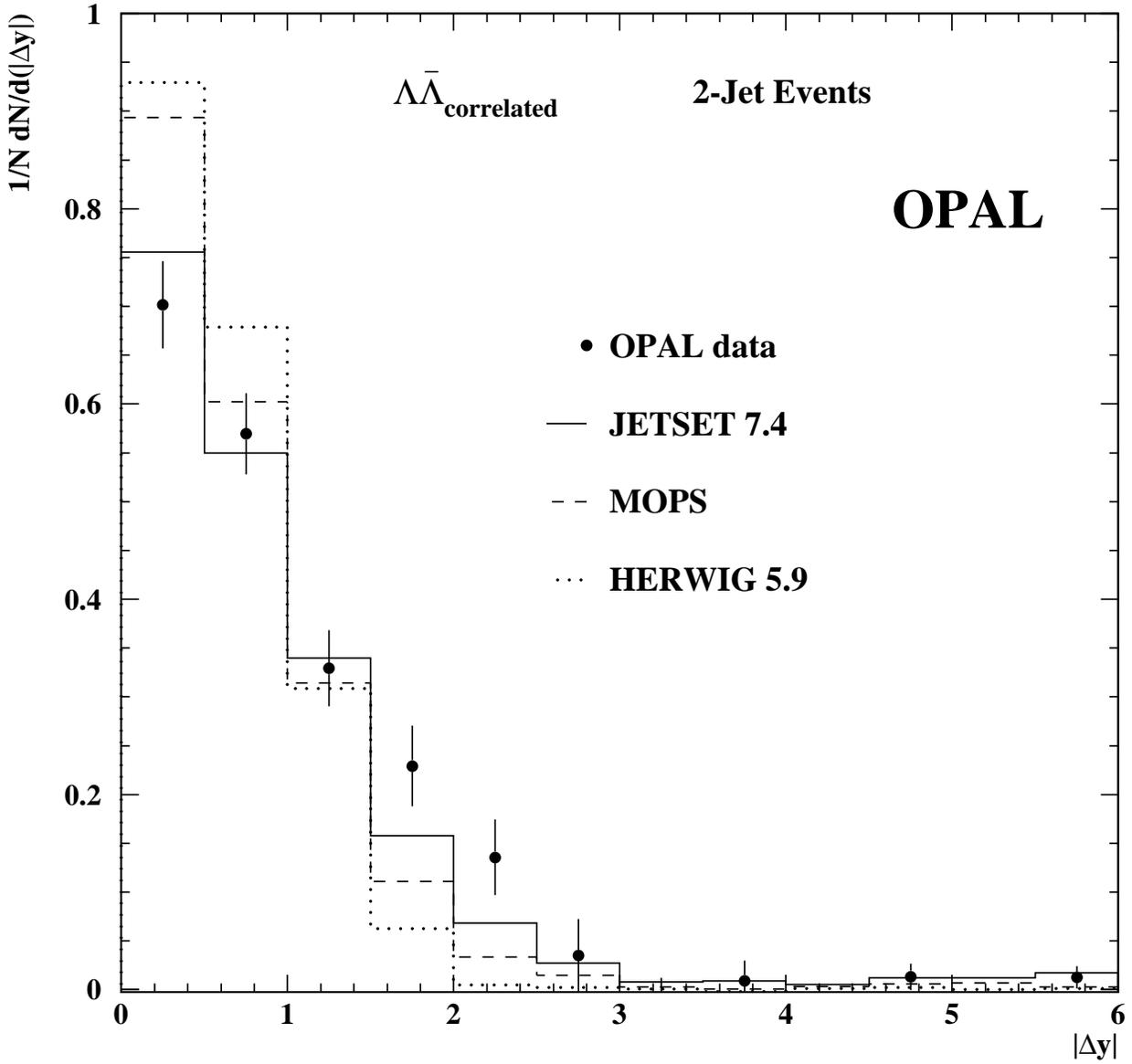}
      }
  \end{center}
  \vspace{-0.5cm}
  \caption{Comparison of the rapidity difference distribution of correlated 
    \llb\ pairs from the 2-jet events with the model predictions.The errors 
    shown are purely statistical, the influence of the systematic errors 
    is negligible.}
  \label{fig:compall_2j}
  \vspace{0.3cm}
\end{figure}

On the other hand, especially in 2-jet events, the rapidity difference \dy\ is 
more sensitive to show differences in the strength of the correlations.
The experimental data and model predictions are compared in 
figure~\ref{fig:compall_2j}.
Again \jt\ gives the best, albeit not completely satisfactory, description of 
the measured distribution. \hw\ generates correlations which are far too 
strong .
The \mops\ model with its built-in facility to allow for several ``popcorn 
mesons'' should yield weaker correlations than \jt ; however, in contrast to 
this naive expectation it produces a narrower \dy\ distribution, i.e. 
stronger correlations. 
We see the following possible reasons for this: first of all, and different 
from \jt , a new kinematic property is built into \mops: the 
low-$\Gamma$-suppression\cite{mops}. This suppresses popcorn fluctuations 
at early times in the color field, resulting in very strong correlations. 
Secondly, it appears that the strength of the correlations is influenced more
by the rate of baryon production via the popcorn mechanism than by the actual 
number of intermediate mesons produced.  
As \jt\ and \mops\ are tuned to show the same mean number of popcorn mesons 
instead of popcorn systems, \mops\ has fewer popcorn systems and therefore 
stronger correlations.

\subsection{Tuning of Models}
\label{subs:tuning}

\begin{figure}[!htb]
  \vspace{-0.7cm}
  \begin{center}
    \resizebox{\textwidth}{!}{
      \includegraphics{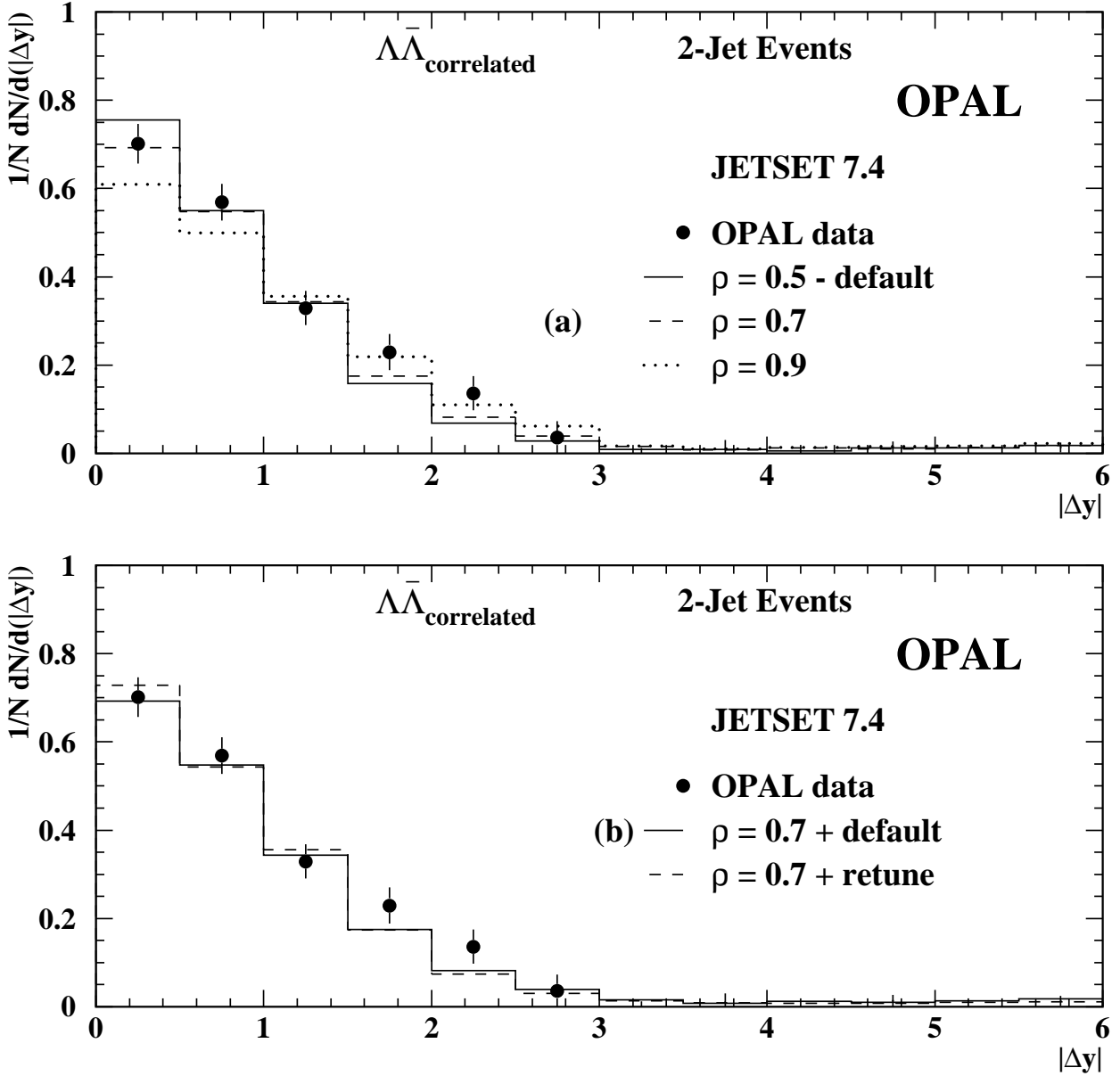}
      }
  \end{center}
  \vspace{-0.5cm}
  \caption{Comparison of the measured rapidity difference distribution of 
    correlated \llb\ pairs in 2-jet events with the \jt\ predictions.  
    (a) Different values of the popcorn parameter (while the other parameters 
    remain at their default values)
    (b) The best popcorn value (0.7) with and without a retune of PARJ(3,4).
    The errors shown are purely statistical, the influence of the 
    systematic errors is negligible.}
  \label{fig:compjt_tunes}
\end{figure}

In an earlier \opal\ analysis of strange baryons\cite{james_early} it was 
observed that the agreement between experimental data and \jt\ predictions
can be improved by adjusting some of the diquark parameters: 
improving the predicted shape of the \dy\ distribution was possible by varying 
the popcorn parameter, $\rho$=PARJ(5), that influences the frequency of 
popcorn production and hence the correlation strength. It was found that 
$\rho$ acts on both the shape of the rapidity spectrum and the production 
rates.
Two other parameters were used to correct for this change of predicted 
multiplicities:
the ratio of the strange to non-strange diquarks over strange to 
non-strange quarks, (us:ud/s:d) = PARJ(3), and the ratio of spin-1 to spin-0 
diquarks, $(1/3 \cdot [{\rm qq}]_1/[{\rm qq}]_0)$=PARJ(4). 
These last parameters affect mainly the rates and leave the spectra nearly
unmodified.
When attempting to improve the predictions of the \mops\ model in the same 
manner, the most direct correspondence to $\rho$ in \jt\ is the 
\mops\ parameter PARJ(8)=$\beta({\rm u})$, the transverse mass of an 
intermediate u-quark. The higher the transverse mass of the intermediate 
system (with several quark pairs possible), the lower the probability to 
produce this popcorn system and the stronger the correlations. 
Therefore, in both \jt\ and \mops\ we tried first to improve the 
agreement with the data distributions by tuning 
the parameters that influence 
the correlation strength (figure~\ref{fig:compjt_tunes} for \jt .) 
In \jt , the popcorn probability $\rho$ was varied from 0\%-90\%, while in 
\mops\ the transverse mass of a u-quark, $\beta({\rm u})$=PARJ(8), was altered 
between 0.2 and 1.0~GeV$^{-1}$. 
All other parameters remained at the \opal\ default values.
As can be seen from table~\ref{tab:jtrates_tune}, for the $\rho$ parameter,
these variations affect not only the shape of the \dy\ distribution but 
also the di-lambda rates, as expected.

The predictions with the different popcorn parameter values in \jt\ are 
compared to the data in figure~\ref{fig:compjt_tunes}a.
Only the results from parameter settings above the default value 
of $\rho = 0.5$ are shown, since lower values give a poorer agreement with the 
data. 
The best agreement 
is found in the range $0.6<\rho<0.8$.
Popcorn values within this range also yield good agreement between data and 
predictions for the \costhet\ distribution.
However, when the influence of the popcorn parameter on the predicted
di-lambda rates is also considered (table~\ref{tab:jtrates_tune})  
use has to be  made of the other two \jt\ parameters that affect the strange 
baryon production in order to tune the rates back to values corresponding to
the measurement. It can be seen from table~\ref{tab:jtrates_tune} and 
figure~\ref{fig:compjt_tunes}b that such
a tune clearly produces a better agreement with the rates (also for single 
particle production) while it does not change the spectra of rapidity 
differences significantly. Using the results of other \opal\ analyses,
it can also be seen that the tune does not change the strange meson (\kshort)
rate, nor does it affect the non-strange baryon (p) rate significantly.
The known problems~\cite{james_early} in modeling the decuplet baryon rates 
are also seen here.
No further attempt has been made to globally optimize the parameter set, 
however. 

The tune of parameter PARJ(8) in \mops\ did not result in an improvement. 
Although the value of the parameter was varied in a comparatively wide 
range, the effect on the \dy\ distribution was almost imperceptible. 
PARJ(8) clearly is not suited to adjust the \mops\ model to the data. 
Therefore, we tested another parameter of the model using the relative 
difference between the fragmentation function $f(z)$ for baryons and mesons, 
the parameter PARJ(45). Again, the variation did not notably change the shape 
of the distribution. 
This relatively poor performance of the \mops\ Monte Carlo in describing 
the \dy\ dependent \llb\ correlations seems to be connected to the known 
shortcomings of the model in describing  $p_{\perp}$-related 
distributions~\cite{mops}.
%
%
%\input{jets}
%
%... corrpaper ... JETS.TEX   /  update: 09-JUL-98
%
\section{Di-lambdas in Jets}
\label{sec:jets}
After studying the strength of \llb\ correlations in the \dy\ spectra, we will 
now present results on the range of the correlations by assigning 
both partners from a correlated pair to the reconstructed jets in an event. 
For short-range correlations both partners are expected 
within the same jet whereas long-range correlations (which can be obtained by 
the production of baryons from the primary quarks) should result in an 
assignment to different jets. 
We use the following two classifications for the assignment study: 
both partners within the same jet, and each partner in a different jet. 

Due to the fact that it is impossible to map 2-jet events at detector level 
to 2-jet events at generator level, we do not attempt to apply efficiency 
corrections but compare our uncorrected results with the \jt\ predictions at 
detector level.
We count the number of \llb\ and \llbb\ pairs in 
each sample and obtain the number of correlated pairs again from the 
relation 
\mbox{$N_{\llb}^{\rm correlated} = N_{\llb} - (N_{\ll} + N_{\lblb})$}.
The amount of background in like- and unlike-sign pairs approximately cancels 
out in this subtraction as long as the contribution from the correlated 
background (see section 3.2) can be neglected. 
The numbers of pairs obtained from the same jet and from different jets are 
listed in table~\ref{tab:injets} 
for both 2- and 3-jet events.
The major part of the correlated pairs is reconstructed 
within the 
same jet (about 96\% in 2-jet events, 81\% in 3-jet events) whereas only a 
very small fraction is found in different jets. These experimental numbers are 
in excellent agreement with the \jt\ predictions at detector level and support
the assumption of short-range compensation of baryon number 
and strangeness in the fragmentation process.
%
%
%\input{summary}
%
%... CORRPAPER ... summary.tex   /  update: 08-JUL-98 RES
%
\section{Summary}
\label{sec:summary}
\llb\ correlations have been studied in 4.3 million multihadronic \zzero\
decays, with the correlated sample obtained from the difference: 
\llbcorr =\llb--(\ll+\lblb).
The analysis has been performed in terms of \costhet\ and rapidity differences 
\dy . As the rapidity is defined with respect to the event (thrust) 
axis, the sensitivity of the analysis is seen to be higher in 2-jet events.
Therefore three data samples have been analyzed: the 
entire hadronic event sample, 2-jet events ($45\%$), and 3-jet events ($36\%$).
The experimental findings have been used to study the baryon production
mechanism implemented in various phenomenological fragmentation models.

The following results have been obtained:
\begin{itemize}
\item  In the full data set, the measured production rates of \llb, \llbb\ 
  and, consequently, \llbcorr\ are in good agreement with a previous \opal\ 
  measurement and results from \aleph\ and \delphi , but show significantly 
  smaller errors. 
\item The \costhet\ distribution of correlated \l-pairs is well suited to 
  distinguish between isotropic cluster and non-isotropic string decay, and 
  clearly favours the latter, implemented in \jt . The predictions of the 
  isotropic cluster model \hw\ are ruled out by the data: they do not describe 
  the features of correlated \llb\ production. 
\item  The rapidity difference \dy\ is used to study the strength of correlated 
  di-lambda production. The measured distribution 
  exhibits strong local correlations.
\item  Satisfactory reproduction of the experimental results is obtained
  with the predictions of the string fragmentation model \jt . Improved
  agreement can be found by tuning some of the default parameters used 
  by \opal . After adjusting the popcorn parameter, to improve the description 
  of the \dy\ spectrum, other parameters, fixing the fraction of diquarks with 
  strangeness and spin1, have to be modified to readjust the predicted rates 
  to the experimental ones. This procedure does not affect the previously 
  optimized \dy\ distribution. 
  The \hw\ model cannot describe the measured \dy\ spectra, and the predictions 
  of \mops , a recently published modification of the \jt\ model, also fail to 
  reproduce the experimental data, even after some parameter tuning. 
\item  In the 2-jet and 3-jet event samples it is found that correlated \llb\ 
  pairs are produced predominantly within the same jet, supporting the 
  assumption of a short-range compensation of quantum numbers.
  Again, the \jt\ predictions are in good agreement with the experimental 
  results. 
\end{itemize} 
In conclusion, the analysis of correlated di-lambda pairs proves to be a
very effective tool to test fragmentation models. \jt\ is the only candidate 
model studied, describing the data successfully. 

\bigskip \bigskip \bigskip \bigskip
%\noindent
%{\bf Acknowledgments} \\
%(to be replaced at time of submission)
\appendix
\par
Acknowledgements:
\par

We are grateful to P. Eden for various interesting discussions about the
performance of the \mops\ model and for his interest in our results. \\
We particularly wish to thank the SL Division for the efficient operation
of the LEP accelerator at all energies
 and for their continuing close cooperation with
our experimental group.  We thank our colleagues from CEA, DAPNIA/SPP,
CE-Saclay for their efforts over the years on the time-of-flight and trigger
systems which we continue to use.  In addition to the support staff at our own
institutions we are pleased to acknowledge the  \\
Department of Energy, USA, \\
National Science Foundation, USA, \\
Particle Physics and Astronomy Research Council, UK, \\
Natural Sciences and Engineering Research Council, Canada, \\
Israel Science Foundation, administered by the Israel
Academy of Science and Humanities, \\
Minerva Gesellschaft, \\
Benoziyo Center for High Energy Physics,\\
Japanese Ministry of Education, Science and Culture (the
Monbusho) and a grant under the Monbusho International
Science Research Program,\\
German Israeli Bi-national Science Foundation (GIF), \\
Bundesministerium f\"ur Bildung, Wissenschaft,
Forschung und Technologie, Germany, \\
National Research Council of Canada, \\
Research Corporation, USA,\\
Hungarian Foundation for Scientific Research, OTKA T-016660, 
T023793 and OTKA F-023259.\\
\newpage
%
%%%%%%%%%%%%%%%%%%%%%%%%%%%%%%%%%%%%%%%%%%%%%%%%%%%%%%%%%%%%%%%%%%%%%%%%%%
% The References                                                         %
%%%%%%%%%%%%%%%%%%%%%%%%%%%%%%%%%%%%%%%%%%%%%%%%%%%%%%%%%%%%%%%%%%%%%%%%%%
%
%\input{references}
%
%... CORRPAPER ... references.tex   /  update: 20.05.98 BN
%

\clearpage
%
%%%%%%%%%%%%%%%%%%%%%%%%%%%%%%%%%%%%%%%%%%%%%%%%%%%%%%%%%%%%%%%%%%%%%%%%%%%
% The tables:
%%%%%%%%%%%%%%%%%%%%%%%%%%%%%%%%%%%%%%%%%%%%%%%%%%%%%%%%%%%%%%%%%%%%%%%%%%%
%
%... CORRPAPER ... tables.tex   /  update: 03.03.98
%
\section*{Tables} 
%
% table 1
%
\begin{table}[!htb]
  \vspace{0.5cm}
  \begin{center}
    \renewcommand{\arraystretch}{1.3}
    \begin{tabular}{|l|l@{ $\pm$ }l@{ $\pm$ }l|l@{ $\pm$ }l@{ $\pm$ }l|
        l@{ $\pm $}l@{ $\pm$ }l|}
      \hline  
      &\multicolumn{9}{c|}{$N_{\rm pairs}$/Hadronic Event [ $\times 10^{-2}$ ]}
\\
      \cline{2-10}
      & \multicolumn{3}{c|}{\llb} 
      & \multicolumn{3}{c|}{\llbb} & \multicolumn{3}{c|}{\llbcorr} \\
      \hline \hline
      This Analysis & 8.95&0.15&0.31  & 2.83&0.11&0.17  
      & 6.12&0.19&0.28 \\
      \hline \hline
      \opal~\cite{opal_corr} & 8.26&0.42&0.79 & 2.05&0.39
      &0.28  & 6.21&0.54&0.84 \\
      \hline
      \aleph~\cite{aleph_corr} 
      &\multicolumn{3}{c|}{9.3 $\pm$ 0.9} 
      &\multicolumn{3}{c|}{2.8 $\pm$ 0.3}    
      &\multicolumn{3}{c|}{6.5 $\pm$ 1.0} \\ 
      \hline
      \delphi~\cite{delphi_corr} 
      & \multicolumn{3}{c|}{9.0 $\pm$ 0.9}  
      & \multicolumn{3}{c|}{1.8 $\pm$ 0.6}    
      &\multicolumn{3}{c|}{7.2 $\pm$ 1.1} \\
      \hline \hline
      \jt\ 7.4 & \multicolumn{3}{c|}{7.75} 
      & \multicolumn{3}{c|}{2.24} & \multicolumn{3}{c|}{5.51} \\
      \hline
      \mops & \multicolumn{3}{c|}{10.57} 
      & \multicolumn{3}{c|}{2.63} & \multicolumn{3}{c|}{7.94} \\
      \hline
      \hw\ 5.9 & \multicolumn{3}{c|}{15.09} 
      & \multicolumn{3}{c|}{3.06} & \multicolumn{3}{c|}{12.03} \\
      \hline
    \end{tabular}
  \end{center}
  \vspace{-0.1cm}
  \caption{ Comparison of average \l\ pair multiplicities from this analysis 
    with those from a previous \opal\ analysis, with the results from other 
    \lep\ experiments and with model predictions. 
    The statistical error is given first, the systematic error second. 
    For \aleph\ and \delphi\ only the total error is available.}
  \label{tab:rates}
  \vspace{0.1cm}
\end{table}
%
% table 2
%
\begin{table}[!htb]
  \begin{center}
    \renewcommand{\arraystretch}{1.3}
    \begin{tabular}{|l|l@{ $\pm$ }l@{ $\pm$ }l|l@{ $\pm$ }l@{ $\pm$ }l|
        l@{ $\pm$ }l@{ $\pm$ }l|} 
      \hline \hline  
      & \multicolumn{3}{c|}{\llb} 
      & \multicolumn{3}{c|}{\llbb} & \multicolumn{3}{c|}{\llbcorr}  \\
      \hline \hline
      & \multicolumn{9}{c|}{$N_{\rm pairs}$/2-Jet Event [ $\times 10^{-2}$ ]}\\
      \hline
      \opal\ data & 5.99 & 0.21 & 0.30 & 1.44 & 0.14 & 0.15 
      & 4.55 & 0.25 & 0.31 \\
      \hline
      \jt\ 7.4  & \multicolumn{3}{c|}{6.14}
      & \multicolumn{3}{c|}{1.45} &\multicolumn{3}{c|}{4.69} \\
      \hline
      \mops & \multicolumn{3}{c|}{8.34} 
      &\multicolumn{3}{c|}{1.68} &\multicolumn{3}{c|}{6.66} \\
      \hline
      \hw\ 5.9 & \multicolumn{3}{c|}{13.16} 
      &\multicolumn{3}{c|}{2.45} &\multicolumn{3}{c|}{10.71} \\ 
      \hline \hline
      & \multicolumn{9}{c|}{$N_{\rm pairs}$/3-Jet Event [ $\times 10^{-2}$ ]}\\
      \hline
      \opal\ data & 9.55 & 0.24 & 0.41 & 2.98 & 0.18 & 0.22 
      & 6.67 & 0.30 & 0.32 \\
      \hline
      \jt\ 7.4 &\multicolumn{3}{c|}{8.70} 
      & \multicolumn{3}{c|}{2.66} & \multicolumn{3}{c|}{6.04}\\
      \hline
      \mops & \multicolumn{3}{c|}{11.92} 
      & \multicolumn{3}{c|}{3.17} & \multicolumn{3}{c|}{8.75}\\
      \hline
      \hw\ 5.9 & \multicolumn{3}{c|}{16.11} 
      & \multicolumn{3}{c|}{3.31} & \multicolumn{3}{c|}{12.80}\\
      \hline \hline
    \end{tabular}
  \end{center}
  \caption{Average multiplicity of \l\ pairs in 2- and 3-jet events  
    compared to model predictions.
    The statistical error is given first, the systematic error second.}
  \label{tab:2_3jrates}
\end{table}

% table 3
\begin{table}[!htb]
  \vspace{0.3cm}
  \begin{center}
    \renewcommand{\arraystretch}{1.3}
    \begin{tabular}{|c|c|c|c|} \hline
      \rule{0mm}{4mm}   & \multicolumn{3}{c|}{Effect on the \llbcorr\ Rate} \\ 
      \cline{2-4}
      \raisebox{1.5ex}[-1.5ex]{Source of Error} & All Hadr. & 2-Jets & 3-Jets \\
      \hline \hline
      Background Systematics     & 0.9\% & 2.8\% & 0.9\%   \\ 
      \hline
      \jt\ 7.3/7.4 Mixing        & 2.3\% & 3.5\% & 1.1\%   \\ 
      \hline
      Monte Carlo Statistics    & 2.3\% & 4.3\% & 3.5\%   \\ 
      \hline
      Cut Simulation             & 3.0\% & 3.0\% & 3.0\%   \\ 
      \hline\hline
      Total Syst. Error          & 4.5\% & 6.8\% & 4.8\%   \\ 
      \hline
      Stat. Error                & 3.1\% & 5.5\% & 4.5\%   \\
      \hline
    \end{tabular}
  \end{center}    
  \vspace{0.3cm}
  \caption{Relative errors in measuring the multiplicity of correlated \l\ 
    pairs in the three event samples.}
  \label{tab:syserr}
\end{table}

% table 4
\begin{table}[!htb]
  \begin{center}
    \renewcommand{\arraystretch}{1.3}
    \begin{tabular}{|c|c|c|c||c||c|} \hline\hline
      &  &  &  & \multicolumn{2}{|c|}{Fraction of \llbcorr} \\ 
      \raisebox{1.5ex}[-1.5ex]{Assignment} & \raisebox{1.5ex}[-1.5ex]{\llb}
      & \raisebox{1.5ex}[-1.5ex]{\llbb} 
      & \raisebox{1.5ex}[-1.5ex]{\llbcorr}
      & \opal\  data & \jt\ det. level \\ \hline\hline
      \multicolumn{6}{|c|}{2-Jet Events} \\ \hline
      Same &  &  &  &  &  \\
      Jet & \raisebox{1.5ex}[-1.5ex]{1994} & \raisebox{1.5ex}[-1.5ex]{469}
      & \raisebox{1.5ex}[-1.5ex]{1525} 
      & \raisebox{1.5ex}[-1.5ex]{(95.6 $\pm$ 2.3)\%}
      & \raisebox{1.5ex}[-1.5ex]{(95.8 $\pm$ 1.8)\%} \\
      \hline
      Different  &  &  &  &  &  \\
      Jets & \raisebox{1.5ex}[-1.5ex]{719} & \raisebox{1.5ex}[-1.5ex]{649} 
      & \raisebox{1.5ex}[-1.5ex]{70} 
      & \raisebox{1.5ex}[-1.5ex]{(4.4 $\pm$ 2.3)\%}
      & \raisebox{1.5ex}[-1.5ex]{(4.2 $\pm$ 1.8)\%} \\
      \hline \hline
      \multicolumn{6}{|c|}{ 3-Jet Events} \\ \hline
      Same &  &  &  &  &  \\
      Jet & \raisebox{1.5ex}[-1.5ex]{2088} & \raisebox{1.5ex}[-1.5ex]{409}
      & \raisebox{1.5ex}[-1.5ex]{1679} 
      & \raisebox{1.5ex}[-1.5ex]{(80.6 $\pm$ 1.8)\%}
      & \raisebox{1.5ex}[-1.5ex]{(80.6 $\pm$ 1.4)\%}  \\
      \hline
      Different &  &  &  &  &  \\
      Jets & \raisebox{1.5ex}[-1.5ex]{1174} & \raisebox{1.5ex}[-1.5ex]{769} 
      & \raisebox{1.5ex}[-1.5ex]{405} 
      & \raisebox{1.5ex}[-1.5ex]{(19.4 $\pm$ 1.8)\%}
      & \raisebox{1.5ex}[-1.5ex]{(19.4 $\pm$ 1.4)\%}  \\
      \hline \hline
    \end{tabular}
  \end{center}
  \caption{Assignment of \l\ pairs to the reconstructed jets in 2-jet 
    and 3-jet events, compared to the predictions of \jt\ 7.4.
    The errors are statistical only.}
  \label{tab:injets}
\end{table}

% table 5
\begin{table}[!htb]
  \begin{center}
    \renewcommand{\arraystretch}{1.3}
    \begin{tabular}{|c|l|l|l||l@{ $\pm$ }l|}
      \hline
      & \multicolumn{3}{c||}{\jt\ 7.4} & \multicolumn{2}{|c|}{ }\\
      \cline{1-4}
      $\rho$  & 0.5* & 0.7 & 0.7 & \multicolumn{2}{|c|}{ } \\
      \cline{1-4} 
      PARJ(3) & 0.45* & 0.45* & 0.60 
      & \multicolumn{2}{|c|}{\raisebox{1.5ex}[-1.5ex]{Data}} \\
      \cline{1-4}
      PARJ(4) & 0.025* & 0.025* & 0.010 & \multicolumn{2}{|c|}{ } \\
      \hline \hline
      \multicolumn{6}{|c|}{Di-lambda Pairs in the Total Hadronic Sample} \\
      \hline
      \llb    & 0.0775 & 0.0668 & 0.0859 & 0.0895 & 0.0034 \\
      \hline
      \llbb   & 0.0224 & 0.0187 & 0.0256 & 0.0283 & 0.0020 \\
      \hline
      \llbcorr & 0.0551 & 0.0481 & 0.0603 & 0.0612 & 0.0034 \\
      \hline \hline
      \multicolumn{6}{|c|}{Di-lambda Pairs in 2-Jet Events} \\
      \hline
      \llb   & 0.0614 & 0.0518 & 0.0677  & 0.0599 & 0.0037 \\
      \hline
      \llbb  & 0.0145 & 0.0119 & 0.0164  & 0.0144 & 0.0021 \\
      \hline
      \llbcorr & 0.0469 & 0.0399 & 0.0513 & 0.0455 & 0.0040 \\
      \hline \hline \hline
      Particle & \multicolumn{5}{c|}{Multiplicities}\\
      \hline
      K$^0$   & 2.02 & 2.03 & 2.02 & 1.99 & 0.04 \\
      \hline
      proton  & 0.93 & 0.89 & 0.92 & 0.92 & 0.11 \\
      \hline
      \l      & 0.338 & 0.316 & 0.361 & 0.374 & 0.010 \\
      \hline
      $\Sigma^+$ & 0.075 & 0.067 & 0.087 & 0.099 & 0.015 \\
      \hline
      $\Sigma^0$ & 0.073 & 0.065 & 0.086 & 0.071 & 0.018 \\
      \hline
      $\Sigma^-$ & 0.068 & 0.059 & 0.080 & 0.083 & 0.011 \\
      \hline
      $\Xi^-$ & 0.0278 & 0.0241 & 0.0341 & 0.0259 & 0.0011 \\
      \hline\hline
      $\Delta^{++}$ & 0.10 & 0.12 & 0.08 & 0.22 & 0.06 \\
      \hline
      $\Sigma(1385)^{\pm}$ & 0.0457 & 0.0546 & 0.0393 & 0.0479 & 0.0044 \\
      \hline
      $\Xi(1530)^0$ & 0.0036 & 0.0040 & 0.0035 & 0.0068 & 0.0007 \\
      \hline
      $\Omega^-$ & 0.0006 & 0.0006 & 0.0006 & 0.0018 & 0.0004 \\
      \hline 
    \end{tabular}
  \end{center}
  \caption{Comparison of inclusive di-lambda yields  with \jt\  Monte Carlo 
    predictions using the \opal\ default tune (second column), a tune to obtain 
    agreement with the measured \dy\ spectra (third column) and a tune to 
    obtain simultaneous agreement in distributions and rates (fourth column).
    The measured values with total errors are given in the fifth column. 
    The single particle inclusive rates are given for comparison: the 
    experimental numbers for \kshort, protons, $\Sigma$ and $\Delta^{++}$ 
    baryons are taken from \cite{k0s}, \cite{protons}, \cite{sigmas}, 
    \cite{delta}, respectively; the remaining numbers are from \cite{opal_sp}.
    The parameters used for the tune are described in the text. 
    Parameter default values are marked with a star.}
  \label{tab:jtrates_tune}
\end{table}
%
%%%%%%%%%%%%%%%%%%%%%%%%%%%%%%%%%%%%%%%%%%%%%%%%%%%%%%%%%%%%%%%%%%%%%%%%%%%
%
%   ..........  EPS-files: ............
%
%%%%%%%%%%%%%%%%%%%%%%%%%%%%%%%%%%%%%%%%%%%%%%%%%%%%%%%%%%%%%%%%%%%%%%%%%%%
\clearpage
%\input{figures.tex}
%
%
%... CORRPAPER ... figures.tex   /  update: 08-JUL-98 RS
%
\section*{Figures}

% figure 1

% figure 2

% figure 3

% figure 4  - entweder die folgenden fig.6-8 verwenden oder stattdessen fig.9

%figure 5 
%
%%%%%%%%%%%%%%%%%%%%%%%%%%%%%%%%%%%%%%%%%%%%%%%%%%%%%%%%%%%%%%%%%%%%%%%%%%%%%
\end{document}